# Novel approach to Room Temperature Superconductivity problem

*Ivan Timokhin[#], Artem Mishchenko[#]*

#HTSC, #betterathome, #iworkfromhome, #DIY

**Abstract |** A long-standing problem of observing Room Temperature Superconductivity is finally solved by a novel approach. Instead of increasing the critical temperature $T_c$ of a superconductor, the temperature of the room was decreased to an appropriate $T_c$ value. We consider this approach more promising for obtaining a large number of materials possessing Room Temperature Superconductivity in the near future.

**Introduction |** The quest of Room Temperature Superconductivity (RTS) exists for a long time, but recently it turned into a race towards RTS, stimulated by progress in rare-earth superhydrides, such as $LaH_{10}$, where the highest critical temperature $T_c$ was observed [1]. Unfortunately, these superhydrides only exist at extreme pressures exceeding hundreds of gigapascals achievable under diamond anvils [2], which limits their practical applications. Furthermore, diamond anvil cells pose difficulties in measuring the Meissner effect, which is an important criterion of observation of superconductivity. Here we propose an alternative approach where RTS is achieved by reducing the temperature of the room. Our approach allows for a clear demonstration of the Meissner effect via a magnetic levitation and flux pinning. For demonstration, we chose a single-crystalline pellet of yttrium barium copper oxide (YBCO), which is a type II superconductor with a relatively high $T_c$ of about 92 K, convenient for our experimental setup. In addition to high $T_c$, YBCO shows strong magnetic levitation, which finds applications in railways, for instance, in Maglev Cobra [3].

**Materials and Methods |** Experimental setup is shown in Fig. 1. It consists of a model room attached to a cold tip of a cryogenic cooler and placed inside a vacuum chamber. The room scale 1:76 (OO gauge) was chosen because of a large number of accessible interiors and figures made for model railways. The room was constructed from a 1/8"-thick rectangular extruded aluminium profile to get a uniform temperature distribution in the room space. The size of the room was also limited by the available cooling power of the cryocooler (5 W at 65 K) and the size of our vacuum chamber. Ricor K535LV cryocooler was purchased on the eBay.com and modified for the air-cooled heat dissipation; vacuum chamber (former load-lock chamber from Nor-Call Inc.) was also obtained from the eBay. For thermal insulation we used multilayer aluminized Mylar sheets cut from party balloons. Interior of the room (back wall with windows and flooring) were printed on Xerox Premium NeverTear paper. The temperature was measured by Cryo-Con Model 12 temperature monitor equipped with Pt100 RTD thermometer attached to the ceiling of the room. We chose a small single-crystal of YBCO as a model superconductor because of a more pronounced quantum locking effect as compared to sintered polycrystalline materials [4].

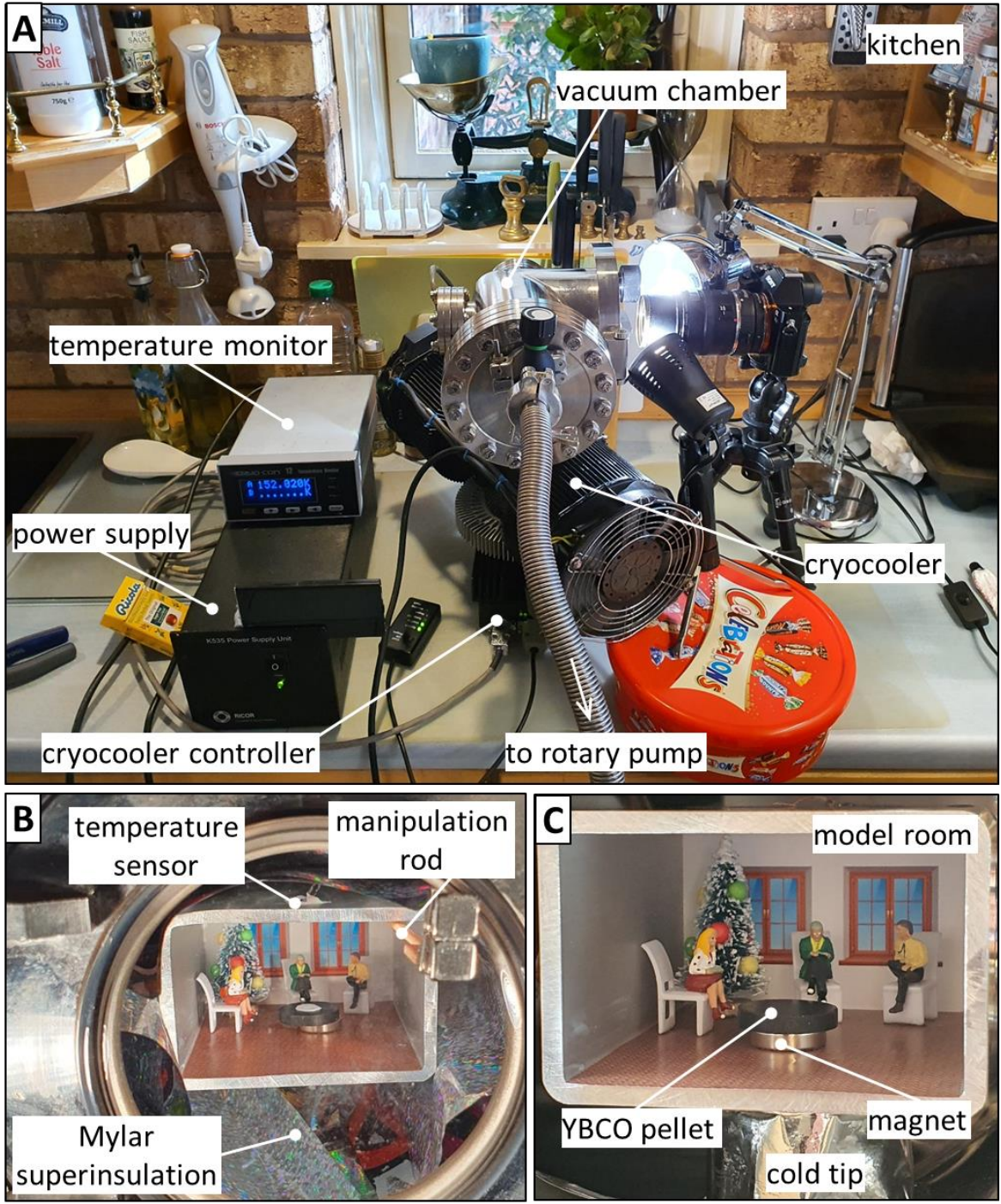

**Figure 1. Photographs of our experimental setup.** (A) Overview of the setup (rotary pump not shown). (B) and (C) Zoom-in photographs of the model room where the quantum locking experiments were performed.

We conducted the experiment as follows: YBCO was attached by a thermal grease to the ceiling of the model room. A strong permanent magnet (N45 grade, diameter 12 mm, thickness 3 mm, magnetic field $B \approx 0.3$ T on the surface) was placed in proximity of the YBCO pellet to achieve a field cooling. The room was evacuated down to $\approx 10^{-3}$ mbar using a rotary pump, then the room was cooled by the cryocooler down to base temperature, $T = 80$ K (at 180 K we disconnected the rotary pump as we found that cryopumping becomes more efficient than the pump).

**Results and Discussion |** Figure 2 highlights the main result of this paper – the observation of quantum locking and Meissner effect in YBCO pellet.

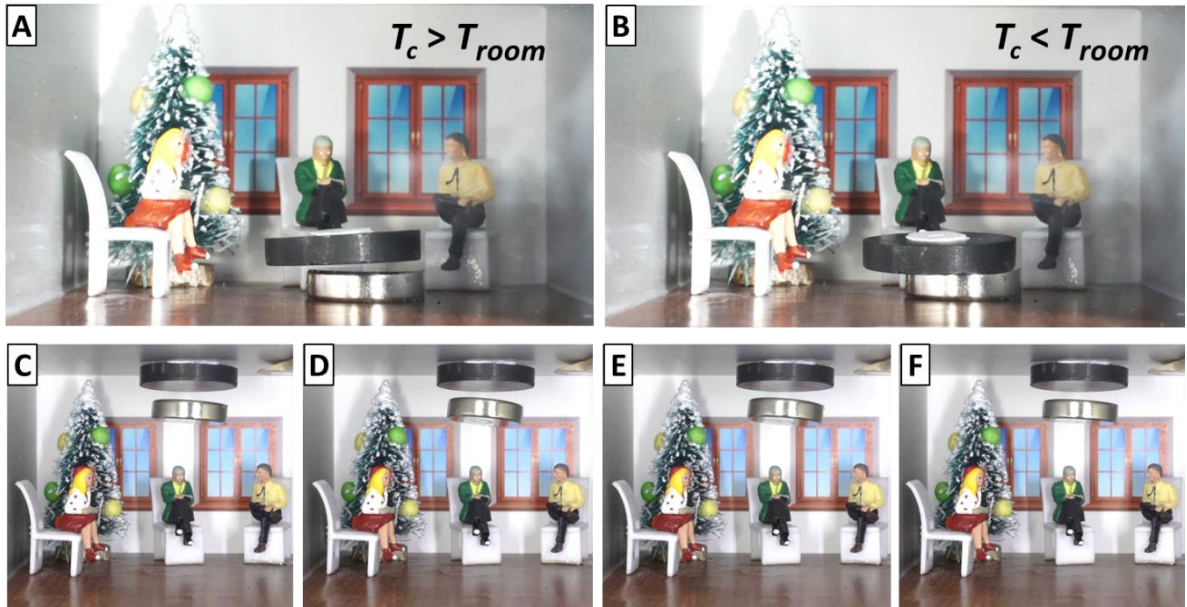

**Figure 2. Observation of quantum locking and Meissner effect in YBCO pellet.** (A) Clear magnetic levitation due to quantum locking is observed when $T_{room}$ is below $T_c$, (B) YBCO pellet falls on the magnet due to gravity force at $T_{room} > T_c$. (C-F) Demonstration of quantum locking in an inverted configuration – the magnet is locked beneath YBCO pellet. (C) The initial position of the pellet and the magnet. (D) and (E) magnet is displaced by applying external (to the room) off-axis magnetic field. (F) Magnet returns to the initial position after the external *B*-field is removed.

When a superconductor above $T_c$ is exposed to a magnetic field, the field lines penetrate freely through its body. During field cooling ($B \neq 0$), a single crystal of type II superconductor transitions to a mixed state below $T_c$, where the field lines concentrate into flux tubes which penetrate through the normal regions of the superconductor surrounded by superconducting quantum vortices [5]. The field cooling results in a quantum locking of the superconductor to the magnet. Figure 2C-F demonstrates the observed quantum locking in a series of photographs: a third, external magnet (outside the chamber, not shown), perturbs magnetic field lines between the YBCO and the locking magnet, resulting in a temporary displacement of the locked magnet, which returns to its original position after the third magnet is moved away. Supplementary video at https://youtu.be/nUYY89lOIjo further demonstrates the quantum locking.

**Outlook |** Our approach shows outstandingly fast development of the new Room Temperature Superconductivity based on the standard YBCO single crystalline material and in future we plan to extend the range of Room Temperature Superconductors to other, more conventional, materials such as $Sn_3Nb$, Pb, Hg etc. by purchasing of a more powerful cryocooler with lower base temperature.

Creating a high-pressure room could also be considered in order to be in line with the modern trends and recent reports about superhydrides. However, Room Pressure Superconductivity paper would be a subject of purchasing ion mill and diamond anvil cell (DAC). Also, a room scale is expected to be much smaller to fit inside the DAC gasket.

**Acknowledgements |** The authors acknowledge unprecedented circumstances of at home lockdown due to COVID-19 outbreak.